\documentclass[letterpaper, 10pt, conference]{ieeeconf}     
 \pdfoutput=1
\makeatletter
\let\NAT@parse\undefined
\makeatother

\usepackage{multirow}
\usepackage{amsfonts}
\usepackage{amsmath}
\usepackage{amssymb}
\usepackage{color}
\usepackage{graphics}
\usepackage{algpseudocode}
\usepackage{caption}
\usepackage{epsfig}
\usepackage{chngcntr}
\usepackage{dashrule}
\usepackage[position=b]{subcaption}
\usepackage[hyperfootnotes=false]{hyperref}
\usepackage{listings}
\usepackage{color}
\usepackage{authblk}
\usepackage{slashbox}

\usepackage{xspace}

\usepackage{amsfonts}
\usepackage{graphicx}

\usepackage{ifthen}
\usepackage{color}

\def\send#1#2{\stackrel{#1}{\hbox to #2{\rightarrowfill}}}
\def\-{\!\!\!\!\!-}

\newtheorem{proposition}{Proposition}

\newcounter{seqn}[equation]
\def\theseqn{\arabic{equation}\alph{seqn}}

\def\endseqn{\eqno \@seqnnum
$$\ignorespaces}
\def\@seqnnum{(\theseqn)}

\newskip\mcentering \mcentering=0pt plus 1000pt minus 1000pt

\def\meqalignno#1{
\halign to\displaywidth{
    \hbox to 0pt{\kern\displaywidth\llap{$##$}\hss}\tabskip=\mcentering
    &\hfil$\displaystyle{##}$\tabskip=\mcentering
   &&$\displaystyle{{}##}$\hfil\tabskip=\mcentering
    \crcr
    #1\crcr}}

\def\dspace{\multiply\normalbaselineskip 150
		  \divide\normalbaselineskip 100 \normalbaselines
		  \csname @@normalbaselineskip\endcsname\normalbaselineskip}
\def\sspace{\multiply\normalbaselineskip 200
		 \divide\normalbaselineskip 300 \normalbaselines
		 \csname @@normalbaselineskip\endcsname\normalbaselineskip}
\def\sdspace{\multiply\normalbaselineskip 160
		 \divide\normalbaselineskip 150 \normalbaselines
		 \csname @@normalbaselineskip\endcsname\normalbaselineskip}

\def\@{\tilde}

\def\3dot#1{\buildrel\textstyle...\over#1}

\newcommand{\Zeta}{Z}
\IEEEoverridecommandlockouts                          

\begin{document}

\author{Yuke Li,\thanks{This work was supported by National Science Foundation grant n.1607101.00 and US Air Force grant n. FA9550-16-1-0290.} Fengjiao Liu, and A. Stephen Morse\thanks{Y. Li is with the Department of Political Science, F. Liu and A. S. Morse are with the Department of Electrical Engineering, Yale University, New Haven, CT, USA, \{yuke.li, fengjiao.liu, as.morse\}@yale.edu}}

\title{A Distributed, Dynamical System View of Finite, Static Games}
\maketitle

\begin{abstract}
This paper contains a reformulation of any $n$-player finite, static game into a framework of distributed, dynamical system based on agents' payoff-based deviations. The reformulation generalizes the method employed in the second part of the study of countries' relation formation problem in \cite{system} to the case of any finite, static game. In the paper two deviation rules are provided and possible applications of this framework are discussed. 

%We therefore propose to study the property of agents' limiting behavior in the dynamical system, instead of the equilibrium points of a static game. We provide two deviation rules, establish the stationary distribution existence of the deviation process, and suggest possibilities for further analysis.


\end{abstract}

\section{Introduction}
\label{intro}

This paper contains a development of a distributed, dynamical system framework using the agents' strategy and payoff information in a finite, static game. In this framework, instead of reaching a static equilibrium, agents alternatively engage in what is termed as long-term ``distributed deviation'' by updating their own strategies myopically and autonomously. An agent can be said to be ``myopic'' if she only focuses on the current payoff rather than the predicted future payoff when she deviates, and ``autonomous'' if she makes the decision only for herself.  The distributed deviation process eventually gives rise to a Markov chain of all strategy profiles in the finite, static game. 

\noindent  {\it Motivating example.} The basic idea of agents' distributed deviation can be first illustrated through a motivating example below, which is the classical ``prisoner dilemma'' game. There are two agents, 1 and 2, each of whom has two possible strategies, ``cooperative'' (C) or ``defective'' (D). Each agent receives a level of payoff associated with each of the four possible combinations of strategies, termed as the four ``strategy profiles''. The table of payoffs is below. 

\begin{center}
  \begin{tabular}{ | l | c | r }
    \hline
    \backslashbox{Agent 2}{Agent 1}
     & (C)& (D) \\ \hline
    (C) & (-1,-1) & (-3, 0) \\ \hline
    (D) & (0,-3) & (-2,-2) \\
    \hline
  \end{tabular}
\end{center}
Two observations can be made regarding the four \emph{strategy profiles} in the prisoner dilemma game from the agents' deviation perspective:

\begin{enumerate}

\item firstly, each agent can only ``locally'' deviate, i.e., within a subset of the four strategy profiles.  For instance, suppose that initially they play $(C,C)$. Individually, agent 1 can only stay at $(C,C)$ or deviate to  $(D,C)$, while agent 2 can only stay at $(C,C)$ or deviate to $(C,D)$.

\item secondly, depending on their deviation order, they may reach quite different strategy profiles. Suppose that only one agent is allowed to deviate from $(C,C)$ at each point in time, and that they myopically deviate. From $(C,C)$, they would expect to arrive at $(D, C)$ or $(C,D)$ depending on which agent is allowed to deviate. And it is possible for them to arrive at $(D,D)$ if both are allowed to deviate at the same time, and have no knowledge of others' deviation. 
 
\end{enumerate}

%As will later be discussed in Section II, we formalize both ``asynchronous deviation'' where at each point in time an agent is allowed to deviate and ``synchronous deviation'' where at each point in time all agents deviate simultaneously. Either deviation finds a list of applications in real life, for which a few are enumerated below. 
%
%The problem of countries' relation formation in \cite{system} has features in terms of both asynchronous deviation and synchronous deviation. Actually, from the perspective of an outsider especially on examining a historic process, agents in many real-life scenarios behave quite myopically, just as in the case of countries' behavior.
%
%About synchronous deviation, an example is trading behavior in the stock market -- traders act without any knowledge of what others might do simultaneously or what might happen in the distant future, and will only (re-)optimize for the next point in time after the outcome of a previous point in time is realized. The same applies to voting, and to many other similar scenarios which may involve a massive amount of agents acting simultaneously (or almost simultaneously) and repeatedly.
%
%Asynchronous deviation is a process featured by the agents who act later in time adjusting to what the agents have acted earlier in time. An example is the migration process in newly found lands, such as the Americas and Australia. In the case of North America, different groups of migrants arrived ``asynchronously'', such as puritans first, more Europeans later, and then the rest of other groups. 

Finite, static games have provided a complete table of all possible strategy combinations of all agents and the associated payoffs for each of them. This is a basis for formulating an alternative problem of distributed deviation by myopic, autonomous agents. There are scenarios or applications in real life that may be best studied as a distributed deviation process of myopic and autonomous agents.  Especially, from the perspective of an outsider with a certain degree of hindsights when examining a historic process such as countries' relation dynamics in \cite{system},  agents in those processes behave myopically, however farsighted they aim to be.  An example is trading behavior in the stock market; traders act without any knowledge of what others might do simultaneously or what might happen in the distant future, and will only act for the next point in time after the outcome of a previous point in time is realized.  Another example is migration processes in new  lands such as the Americas and Australia.  In the case of North America, different groups of migrants arrived ``asynchronously'', with pilgrims and puritans among the first, and the rest to follow.

In Section II, the transitions among strategy profiles made possible by agents' individual deviations will be studied as a distributed, dynamical system using the payoff information from the finite, static game. Each agent \emph{only} needs to know her own payoff information in order to make deviations in the dynamical system. 

%The same applies to voting, and to many other similar scenarios which may involve a massive amount of agents acting simultaneously (or almost simultaneously) and repeatedly.

\noindent  {\it In relation to the existing literature.} This paper generalizes the method in \cite{system} initially developed for offering a dynamical perspective toward the countries' relation formation problem to the case of any $n$-player finite, static game.  The differences of this paper from the approaches in several related strands of literature and, relatedly, some problems where it can most suitably apply are the following: 

First, the paper develops the dynamical system framework drawing on Markov chain. The class of games extensively using Markov decision processes is stochastic games \cite{shapley1953stochastic}. But the major difference with stochastic games,  the dynamics of agents' deviation process in this paper do not require agents to act in mutual best responses as required by game equilibria. 

Second, the framework in this paper relates to the following work on payoff-based learning in games, such as \cite{marden2014achieving,marden2010revisiting,marden2009joint} and \cite{marden2009payoff}, in the sense that agents' distributed deviation in this paper will also be payoff-based. However, in this paper agents are assumed to deviate to where they are better off, not necessarily where is their best responses. 

Third, the framework in this paper can be used for the purpose of equilibrium selection for finite, static games; this is achieved by designing the agents' deviation process, e.g., how they transition among the strategy profiles. Some well-known equilibrium selection methods were developed for evolutionary games (e.g., \cite{kandori1993learning}, \cite{robson1996efficient} and \cite{samuelson1998evolutionary}, and for signaling games (e.g., \cite{banks1987equilibrium}). Equilibrium selection was done based on a certain perfection criteria for equilibria (e.g.,  \cite{harsanyi1988general, myerson1978refinements, carbonell2011existence}) or a certain strategy revision process \cite{blume1993statistical}. It appears that the equilibrium selection method in this paper is different from all these existing ones.

Lastly,  the framework in this paper is technically similar to graph dynamical system (e.g., \cite{kuhlman2011general}), of which two variants are cellular automata and interacting particle systems.  Graph dynamical systems represent a useful approach for capturing distributed, dynamical phenomena with local interactions.  A review of applications of methods in statistical mechanics in social issues can be found in \cite{castellano2009statistical}.  However, in many of these works, the evolution rules are usually not payoff or preference-based (an exception is \cite{meyer1998statistical}), which explains the major difference from this paper. Therefore, this framework may be applied in those issue areas to arrive at possibly different conclusions.

\noindent  {\it Structure of the paper.}  The development of this paper proceeds with the following. Firstly, the distributed, dynamical system framework of agents' distributed deviation will be proposed in Section II. Secondly,  some analysis of the framework will be provided in Section III. Thirdly, possible applications of this framework will be illustrated in Section IV. %, such as studying some new questions of interest.  %Most significantly, we claim that this helps to define a new problem of agents' {\color{red} \emph{optimization of the stationary distribution of strategy profiles}} with distributed deviation.

\section{The Framework} 
\label{deviation}

\subsection{Finite, Static Game.} A finite, static game is generally formulated as follows.  Suppose there are a finite number of agents indexed by the elements in $\mathcal{V} = \{1,2,...,n\}$. Each of them has a finite strategy set $\mathcal{S}_{i}$ and $s_{i} \in \mathcal{S}_{i}$ is a strategy of agent $i$. 

The set of all possible strategy profiles $\mathcal{S}$ in the game-theoretic terminology, i.e., the strategy combination of all agents, is therefore the Cartesian product of all agents' strategy sets $\mathcal{S}_{1} \times \mathcal{S}_{2} \times ... \times \mathcal{S}_{n}$, and a finite set. Let $s = [s_{1}~s_{2}~ \dots ~s_{n}]$ denote a strategy profile in  $\mathcal{S}$. Let $$p: \mathcal{S} \to \mathbb{R}^{n}$$ be the payoff function such that for $s \in \mathcal{S}$, $p(s) = [p_{1}(s)~p_{2}(s) ~\dots ~p_{n}(s)]$ where $p_{i}(s)$ is the payoff for agent $i$ from the strategy profile $s$, for each $i \in \mathcal{V}$. With the components being so defined, a game can therefore be specified as the collection $\Gamma = \{\mathcal{V}, \mathcal{S}, p\}$. 

\subsection{Agents' distributed deviations}
At time $t+1$, each agent updates its own strategy based on the strategies of others at time $t$. It should be noted that it is not required that an agent's payoff is affected by \emph{all} others' strategies.  While updating its own strategy, agent $i$ implicitly assumes that the strategies of all other agents remain the same. For simplicity, let the strategy profile at time $t$ be $u$ and the strategy profile at time $t+1$ be $v \in \mathcal{S}$.

\noindent {\it Set of deviations from a strategy profile.} From $u$, agent $i$ can only deviate to a subset of $\mathcal{S}$, which we define as  agent \emph{$i$'s set of deviations from $u$} and denote using $\mathcal{S}^{i}_{u}$, in which $u$ is by default an element. A strategy profile in $\mathcal{S}^{i}_{u}$ that is not $u$ is only different from $u$ in terms of the $i$-th coordinate, i.e., agent $i$'s strategy. 

In the previous example, from the strategy profile $(C,C)$, agent 1's deviation set is $\{(C,C), (D,C)\}$.

\noindent {\it Individual transition probability.} Any agent chooses a time-invariant nonnegative probability called \emph{individual transition probability} with which to transition from a strategy profile to another. Any agent' transition probabilities are based solely on her own payoff information. Formally speaking, the map $$\psi_{i}: \mathcal{S} \times \mathcal{S} \to [0,1]$$ denotes agent $i$'s transition probability from a strategy profile to another.

If $v$ is not in agent $i$'s deviation set from $u$, we stipulate that $\psi_{i}(u,v) = 0$; otherwise, $\psi_{i}(u,v)$ is a time-invariant function of agent $i$'s payoffs specified in the game $\Gamma$. 

In the previous example, $(D,D)$ is not in agent 1's deviation set from $(C,C)$. Agent 1's total probabilities of staying at $(C,C)$ and transitioning to $(D,C)$ sum to 1.

\noindent {\it Deviation order.}  Two possible deviation orders are:

\begin{enumerate}

\item \emph{Asynchronous updating.} The first deviation rule is where only one agent deviates at each time. Suppose only one agent is picked based on a nonnegative probability $a_{i}$ to deviate at any time instant, where the total probabilities for all agents sum to 1 as below. The received chance for agent $i$ to deviate can be a good proxy of its ``agenda-setting power''.

\begin{equation*}
a_{i} \in [0,1], ~\texttt{and}~ \sum\limits_{i\in \mathcal{V}}a_{i} = 1
\end{equation*}

In the previous example, suppose agent 1 and agent 2 both receives 1/2 probability of deviation. For instance, suppose that agent 1 transitions to $(D,C)$ with probability 1 and agent 2 transitions to $(D,C)$ with probability 0. The transition probability from $(C,C)$ to $(D,C)$ will then be 1/2.

\item \emph{Synchronous updating.} The second deviation rule is where agents deviate simultaneously. In synchronous deviation, any agent deviates by assuming all others' strategies to remain fixed. Therefore, the strategy profile they eventually arrive at can be different from what they expect themselves to deviate to.  This gives us a definition of  ``anticipated strategy profile''. Denote $v^{i}_{u} \in \mathcal{S}_{u}^{i}$ as the anticipated new strategy profile by $i$ when deviating from strategy profile $u$, where \emph{only} its $i$-th coordinate is the same as that of $v$, and the rest of the $n-1$ coordinates are the same as those in $u$.  %Intuitively,  the change from $u$ to $v$ depends on all agents' deviations. 

\end{enumerate}

\noindent {\it Strategy profile transition probability.} Let the transition probability matrix of the strategy profiles be $\Zeta = [\zeta_{uv}] \in [0,1]^{|\mathcal{S} | \times |\mathcal{S}|}$, where $\zeta_{uv}$ represents the \emph{transition probability} from a strategy profile $u$ to another $v$.  The strategy profile transition probability is reflected in the arc weight of $\mathbb{H}$ as previously defined. Time-invariant $\Zeta$ is the focus of this paper.

Next is a discussion of how agents' deviation order (i.e., asynchronous or synchronous) impacts the way in which the individual transition probability from $u$ to $v$, $\psi_{i}(u,v)$, aggregates to the strategy profile transition probability from $u$ to $v$, $\zeta_{uv}$. 

\begin{enumerate}
\item {\it In asynchronous deviation.} The transition probability from $u$ to $v$, $\zeta_{uv}$, can be expressed as 
\begin{equation*}
\zeta_{uv} = \sum\limits_{i\in \mathcal{V}}a_{i}\psi_{i}(u,v). 
\end{equation*}

\item {\it In synchronous deviation.} The transition probability from $u$ to $v$, $\zeta_{uv}$, is the product of every agent's individual transition probability from $u$ to her anticipated strategy profile $v^{i}_{u}$,

\begin{equation*}
\zeta_{uv} =  \prod\limits_{i \in \mathcal{V}} \psi_{i}(u,v^{i}_{u}).
\end{equation*}

In the previous example, suppose agent 1 transitions to $(D,C)$ with probability 1 and agent 2 transitions to $(C,D)$ with probability 1, the transition probability from $(C,C)$ to $(D,D)$ will be 1. In fact, the transition probability from any other strategy profile to $(D,D)$ will be 1, and the probability of staying at $(D,D)$ is also 1. 

\end{enumerate}

\noindent {\it Deviation process.} Below is a discussion of the elements necessary to understand the deviation process. 

\begin{enumerate}
\item {\it Initial distribution.} Agents can be regarded as being initially involved in a distribution of all strategy profiles in $\mathcal{S}$.  

Denote the distribution of all the strategy profiles in $\mathcal{S}$ at time $T$ as $\pi^{T}$. $\pi^{T} = [\pi^{T}_{u}]^{1 \times |\mathcal{S}|}$,  where $\pi^{T}_{u} \in [0,1]$ representing the probability mass of staying in the strategy profile $u$ at time $T$, and $\|\pi^{T}\|_{1} = 1$. Obviously, $\pi^{0}$ represents the \emph{initial distribution}.

\item {\it Deviation process.} The dynamical system is the collection $\Omega = (\Gamma, \pi^{0}, \Zeta\}$. Agents' deviation process can be conveniently understood as applying the operator $\Zeta$ repeatedly to the initial distribution, i.e., $\pi^{0}$. 

The Ces\`{a}ro average as below
\begin{equation*}
\lim\limits_{n\to\infty}\frac{1}{n}\sum\limits_{k=1}^{n}\pi^{0}Z^{k}
\end{equation*} 
represents the limiting distribution of the strategy profiles in the deviation process.

\end{enumerate}

\subsection{Assumptions on agents' transition probabilities} 

An agent's transition probability between strategy profiles can be assumed to satisfy the following basic assumptions: 

\begin{enumerate}
\item The total probabilities of staying at original strategy profiles and transitioning to different ones in either deviation are normalized to be 1.  

In asynchronous updating,\begin{equation*}
\sum\limits_{v \in \mathcal{S}_{u}^{i}}\psi_{i}(u,v) = 1.
\end{equation*} 
In synchronous updating, 
\begin{equation*}
\sum\limits_{v^{i}_{u} \in \mathcal{S}_{u}^{i}}\psi_{i}(u,v^{i}_{u}) = 1
\end{equation*}

\item It can be assumed that in either updating from any strategy profile agents will only assign positive probability to other strategy profiles giving better payoffs; otherwise, they will remain playing the original strategy profile $u$ with probability 1.  Technically speaking, 

If $\not\exists v \in \mathcal{S}_{u}^{i}$ such that $p_{i}(v) > p_{i}(u)$, then 
\begin{equation*}
\psi_{i}(u,v) = 0, \forall v \in \mathcal{S}, v \neq u, {~\text{and}~  \psi_{i}(u,u) = 1}
\end{equation*}
 Otherwise, 
 \begin{equation*}
 \psi_{i}(u,v) > 0
 \end{equation*}

%\item In either updating, we assume that any agent will only deviate within its deviation set, and will not deviate to strategy profiles which give her fewer or equal payoffs. 
%
%If $p_{i}(v)  \leq p_{i}(u)$ or $v \notin \mathcal{S}_{u}^{i}$,  
%\begin{equation*}
%\psi_{i}(u,v) = 0.
%\end{equation*}
%
%
%\item In either updating, we assume that any agent will only deviate with positive probability to strategy profiles which give her better payoffs. If $v \in \mathcal{S}_{u}^{i}$ and $p_{i}(v)  > p_{i}(u)$, 
%\begin{equation*}
%\psi_{i}(u,v) > 0.
%\end{equation*}

\end{enumerate}

\section{Analysis and Results}

\noindent {\it Property of the strategy profile transition graph.} In this section the strategy profile transition graph $\mathbb{H} = \{\mathcal{S}, \mathcal{A}\}$ will be defined and discussed.

A \emph{strategy profile transition graph} $\mathbb{H} = \{\mathcal{S}, \mathcal{A}\}$, is a directed and weighted graph, where each strategy profile (or ``state'') in $\mathcal{S}$ will be regarded as a vertex and any transitions in the strategy profiles as ``walks'' from a vertex to another.  The weight of the arc directed from one vertex to another represents the transition probability from one strategy profile to another.  For $i \in \mathcal{V}$, let $d_i$ be the trivial metric on $\mathcal{S}_i$, i.e., $d_i(u_{i},v_{i})=0$ if $u_i = v_i \in \mathcal{S}_i$ and $d_i(u_{i},v_{i})=1$ if $u_{i} \neq v_{i}$, and $u_{i}, v_{i} \in \mathcal{S}_i$. A metric $d: \mathcal{S} \times \mathcal{S} \to \{0, 1, 2, \dots, n\}$ can be defined on the set $\mathcal{S}$ such that for $u, v \in \mathcal{S}$, $d(u,v)$ is the number of coordinates of $u$ that differ from $v$, i.e.,$$d(u,v) = \sum_{i=1}^n d_i(u_{i}, v_{i}),$$ where $u_i$ and $v_i$ are the $i$-th coordinates of $u$ and $v$, respectively. 

 In the case of asynchronous deviation, $\mathbb{H}$ will \emph{only} contain arcs between strategy profiles with distance at most 1. However, in the case of synchronous deviation, $\mathbb{H}$ may also contain arcs between strategy profiles with distance up to $n$. 

The weights of the arcs depends on individual transition probabilities. An arc in $\mathbb{H}$ with zero weight means that this arc does not exist. As in the previous example, an arc does not exist between $(C,C)$ and $(D,D)$ with distance 2 in asynchronous deviation, while an arc between them does exist in synchronous deviation. But if we adjust the payoffs, the individual transition probabilities will then change accordingly, in which case the arc between $(C,C)$ and $(D,D)$ may no longer exist  in synchronous deviation.

\noindent  {\it Convergence.}  In this section it is proven in Proposition~\ref{prop/sum} that  $\Zeta$ is always a stochastic matrix. %Accordingly, Propositions~\ref{prop/stationary} further establishes that a stationary distribution always exists for $\Zeta$, which will give us predictions regarding the distribution of strategy profiles agents involve in during the process.

\begin{proposition}
\label{prop/sum} In either asynchronous updating or synchronous updating, every row of $\Zeta$ sums to 1. 

\end{proposition}{\bf Proof of Proposition~\ref{prop/sum}:} In asynchronous updating, 

\begin{align*}
\sum\limits_{v \in \mathcal{S}}\zeta_{uv} = \sum\limits_{v \in \mathcal{S}}\sum\limits_{i \in \mathcal{V}}a_{i}\psi_{i}(u,v) 
= \sum\limits_{i \in \mathcal{V}}a_{i}\sum\limits_{v \in \mathcal{S}_{u}^{i}}\psi_{i}(u,v)
\end{align*}

Since $\sum\limits_{v \in \mathcal{S}_{u,i}}\psi_{i}(u,v) = 1$, then 

\begin{align*}
\sum\limits_{v \in \mathcal{S}}\zeta_{uv} = \sum\limits_{i \in \mathcal{V}}a_{i} = 1
\end{align*}

In synchronous updating, 

\begin{align*}
\sum\limits_{v \in \mathcal{S}}\zeta_{uv} = \sum\limits_{v \in \mathcal{S}}\prod\limits_{i \in \mathcal{V}} \psi_{i}(u,v^{i}_{u}) 
\end{align*}

The above expression can be equivalently expressed as

\begin{align*}
\sum\limits_{v \in \mathcal{S}}\prod\limits_{i \in \mathcal{V}} \psi_{i}(u,v^{i}_{u}) = \prod\limits_{i \in \mathcal{V}}\sum\limits_{v \in \mathcal{S}_{u}^{i}} \psi_{i}(u,v) = 1
\end{align*} 

Therefore, under both rules, every row of $\Zeta$ sums to 1. $\square$

An argument for any Markov Chain on finite state space to have a stationary distribution is given by the Brouwer Fixed Point Theorem.  The Ces\`{a}ro average as discussed earlier
\begin{equation*}
\lim\limits_{n\to\infty}\frac{1}{n}\sum\limits_{k=1}^{n}\pi^{0}Z^{k}
\end{equation*} defines the limiting distribution of the strategy profiles, which will be of interest for studying this the deviation process. 

Basically, this framework can be used for the relationship between the game predictions and the dynamical system predictions. There can be some equivalence results between the game predictions and the dynamical system predictions. Two of these results are stated in Proposition 2 and Proposition 3.

\noindent  {\it Linkage with pure strategy Nash equilibrium.} Proposition \ref{prop/eigen} explores a link between the distributed, dynamical system and the original static, finite game, by suggesting that an equilibrium strategy profile in the static game is always a stationary distribution.

Recall the definition that a strategy profile $s^{*}$ is a pure strategy Nash equilibrium if no agent would strictly benefit in terms of payoff by deviating to any other strategy profile $\hat{s} \in \mathcal{S}$,  that is, 

\begin{equation*}
\forall \hat{s} \in \mathcal{S}, p_{i} (\hat{s}) \leq p_{i}(s^{*}), \forall i \in \mathcal{V}
\end{equation*}

\begin{proposition}
\label{prop/eigen}

A Nash equilibrium $s^{*}$ of the static game $\Gamma$ corresponds to a unit left eigenvector of $\Zeta$. Such a left eigenvector is a stationary distribution of $\Zeta$. 

\end{proposition}{\bf Proof of Proposition~\ref{prop/eigen}:} Suppose $s^{*} \in \mathcal{S}$ is the $k$-th strategy profiles. Since $s^{*}$ is a Nash equilibrium, which means that no agent is willing to deviate, and by definition, $\forall s \neq s^{*} \in \mathcal{S}, \forall i \in \mathcal{V}, \psi_{i}(s^{*}, s) = 0$. By decomposability of $\Zeta$, we have
\begin{align*}
\Zeta = \left[\begin{array}{ccc}
A_{(k-1) \times (k-1)} & \alpha_{(k-1) \times 1} & B_{(k-1) \times (n-k)}\\
0_{1 \times (k-1)} & 1 & 0_{1 \times (n-k)}\\
C_{(n-k) \times (k-1)} & \beta_{(n-k) \times 1} & D_{(n-k) \times (n-k)}\\
\end{array}\right]
\end{align*}

Let $e_{k} = [\underbrace{0, \ldots, 0}_{(k-1)~0's}, 1,\underbrace{0, \ldots, 0}_{(n-k)~0's}]$ be the correspondent eigenvector of $\Zeta$. It is easy to know that $e_{k}\Zeta = e_{k}$, which means that the unit vector $e_{k}$ is a left eigenvector of $\Zeta$. Since $\|e_{k}\|_{1} = 1$, it is also a stationary distribution of $\Zeta$.  $\square$

If $\Gamma$ has unique pure strategy Nash equilibrium, $\Zeta$ will have a stationary distribution in which the probability mass of that strategy profile is $1$. If $\Gamma$ has multiple pure strategy equilibria, any vector in the span of correspondent unit eigenvectors of $\Zeta$ is a stationary distribution.

\noindent {\it Linkage with mixed strategy Nash equilibrium.}  For instance,  the cases in which the mixed strategy Nash equilibrium of the finite game coincide with the eigenvector of the transition matrix of the dynamical system can be derived. 

Let $\Gamma$ be a finite game with two agents, in which agent 1 has two strategies to play with, namely $U$ and $D$, and agent 2 also has two strategies to play with, namely $L$ and $R$. The payoff matrix of $\Gamma$ is shown below. For example, if agent 1 plays strategy $U$ and agent 2 plays strategy $L$, the payoffs received by agent 1 and agent 2 are $p_1(U,L)$ and $p_2(U,L)$, respectively. 

Now suppose $p_1(D,L) > p_1(U,L)$ and $p_1(U,R) > p_1(D,R)$ for agent 1; $p_2(U,L) > p_2(U,R)$ and $p_2(D,R) > p_2(D,L)$ for agent 2. Then the game has no pure strategy Nash equilibrium and the strategy profile deviation process follows a counter-clockwise cycle. Of course a mixed strategy Nash equilibrium exists, in which agent 1 plays strategy $U$ with probability $a$ and agent 2 plays strategy $L$ with probability $b$, where $0 < a,~b < 1$.

\begin{table}[!h]
\centering
\caption{The payoff matrix of $\Gamma$}
\label{game1}
\begin{tabular}{|l|l|l|l|}
\hline
\multicolumn{4}{|l|}{\qquad\qquad\qquad\qquad\qquad\qquad\qquad agent 2} \\ \hline
\multirow{5}{*}{agent 1} &  & \qquad\qquad $L$ & \qquad\qquad $R$ \\ \cline{2-4} 
 &   &   & \\[-1em]
 & $U$ & ($p_1(U,L)$, $p_2(U,L)$) & ($p_1(U,R)$, $p_2(U,R)$) \\ \cline{2-4}
 &   &   & \\[-1em]
 & $D$ & ($p_1(D,L)$, $p_2(D,L)$) & ($p_1(D,R)$, $p_2(D,R)$) \\ \hline
\end{tabular}
\end{table}

Then the discrete-time strategy profile deviation process is modeled as a Markov chain. The four states of the Markov chain are the four strategy profiles. At each time instant, assume the two agents will synchronously and myopically choose whether to update their own strategies from the last time instant. A family of the transition matrices for the Markov chain, $Z$, is

\begin{equation*}
\begin{array}{c|cccc}
          & 1.(U,L) & 2.(U,R) & 3.(D,L) & 4.(D,R)  \\ \hline
1.(U,L) & 1-\rho_1  & 0         & \rho_1    & 0          \\
2.(U,R) & \rho_2    & 1-\rho_2  & 0         & 0          \\
3.(D,L) & 0         & 0         & 1-\rho_3  & \rho_3     \\
4.(D,R) & 0         & \rho_4    & 0         & 1-\rho_4
\end{array}
\end{equation*}
where $0 < \rho_1,~\rho_2,~\rho_3,~\rho_4 < 1$.

\begin{proposition} \label{lem1:mix-eq}
If the transition probabilities of the Markov chain satisfy $\rho_1 : \rho_2 : \rho_3 : \rho_4 = (p_1(D,L) - p_1(U,L))(p_2(U,L) - p_2(U,R)):(p_1(U,R) - p_1(D,R))(p_2(U,L) - p_2(U,R)):(p_1(D,L) - p_1(U,L))(p_2(D,R) - p_2(D,L)):(p_1(U,R) - p_1(D,R))(p_2(D,R) - p_2(D,L))$, the unique stationary distribution is $[ab ,\enspace a(1-b) ,\enspace (1-a)b ,\enspace (1-a)(1-b)]$.
\end{proposition}

\noindent \textbf{Proof of Proposition~\ref{lem1:mix-eq}:} Let $\pi = [ab ,\enspace a(1-b) ,\enspace (1-a)b ,\enspace (1-a)(1-b)]$. By the definition of mixed strategy Nash equilibrium, 
%\begin{equation*}
$$a = \frac{p_2(D,R)-p_2(D,L)}{p_2(D,R)-p_2(D,L)+p_2(U,L)-p_2(U,R)}$$
$$b = \frac{p_1(U,R)-p_1(D,R)}{p_1(U,R)-p_1(D,R)+p_1(D,L)-p_1(U,L)}$$
%\end{equation*}
Let 
\begin{align*}
\rho_1 &= (p_1(D,L) - p_1(U,L))(p_2(U,L) - p_2(U,R)) \rho_0,  \\ \rho_2 &= (p_1(U,R) - p_1(D,R))(p_2(U,L) - p_2(U,R)) \rho_0 \\
\rho_3 &= (p_1(D,L) - p_1(U,L))(p_2(D,R) - p_2(D,L)) \rho_0,  \\ \rho_4 &= (p_1(U,R) - p_1(D,R))(p_2(D,R) - p_2(D,L)) \rho_0
\end{align*}
where $\rho_0$ is small enough to ensure $0 < \rho_1,~\rho_2,~\rho_3,~\rho_4 < 1$. It can be checked that $\pi Z = \pi$, so $\pi$ is a stationary distribution. As the Markov chain is irreducible and aperiodic, $\pi$ is the unique stationary distribution. \hfill $\square$ 

In the mixed Nash strategy equilibrium of this game, the probability of the two players playing strategy profiles $(U, L)$, $(U, R)$, $(D, L)$, and $(D, R)$ are $ab$, $a(1-b)$, $(1-a)b$, and $ (1-a)(1-b)$, respectively, which is exactly the stationary distribution of the Markov chain. Proposition~\ref{lem1:mix-eq} also illustrates that by properly choosing the state transition probabilities of the Markov chain based on the payoffs, each player can optimize its own payoff.

%\textcolor{red}{comment: If the equations are too long to fit in the page, we may number the four strategy profile as 1, 2, 3, 4 and use $p_1^4$ for payoff of agent 1 at 4th strategy profile.}

\section{Discussion and Conclusion}

\noindent  {\it Summary of the framework.}

\begin{enumerate}
\item {\it Two graphs.} Other than the strategy profile transition graph, there exists another two graphs from the framework, which may be useful for the analysis with regards to certain variations of finite, static games. 

First, a finite, static game provides the required information for setting up a \emph{neighbor graph} of agents, where an agent's neighbors in this graph are those whose strategies will affect its payoff.

Second, the game has the required information for setting up an \emph{interaction graph} of agents. In an interaction graph, the neighbors of agents are those toward whom agents direct their strategies, and the strategy toward one neighbor is independent of the strategy toward the other. This is especially so if the agents are connected through a network. The difference from a neighbor graph is that an agent can not necessary affect the payoffs of her neighbors in the interaction graph. 

\item {\it Two probabilities.} The framework has two important probabilities, which are agents' individual transition probabilities between the strategy profiles and the probabilities of the strategy profiles in the stationary distribution, with the former affecting the latter and, of course, the weights of the arcs in the strategy profile transition graph. 

\end{enumerate}
%Agents' individual transition probabilities can be treated as , the properties of the strategy profile transition graph, and the stationary distribution of the strategy profiles, together.

\noindent  {\it How to use the framework?} Two possibilities of further using the framework are discussed below.

\subsubsection{From systems to games} Based on the properties of the dynamical system, predictions of interest for the original game itself can be derived. As discussed in the introduction, to apply the framework to select equilibria of the game is an example. As is commonly known, many games have multiple kinds of equilibria and it is impossible ex ante to predict which equilibrium to be played by agents. The equilibria selected can be a mixed strategy Nash equilibrium, a pure strategy Nash equilibrium, or Pareto-optimal Nash equilibrium such as in the example below. 

For instance, let $\Gamma$ be a finite game with two agents, in which agent 1 has three strategies to play with, namely $U$, $C$ and $D$, and agent 2 also has three strategies to play with, namely $L$, $M$ and $R$. The payoff matrix of $\Gamma$ is shown below. In this game, there is a pure strategy Nash equilibrium, which is $(D,R)$, and a mixed strategy Nash equilibrium, which is $((1/2, 1/2, 0), (1/2, 1/2, 0))$, where both agents are involved in the cycle of the 1, 2, 4 and 5-th strategy profiles.

\begin{table}[htbp]
\centering
\caption{A Game Example}
\label{my-label}
\begin{tabular}{|l|l|l|l|l|}
\hline
\multicolumn{5}{|l|}{\qquad\qquad\qquad\qquad agent 2} \\ \hline
\multirow{4}{*}{agent 1} &  & $L$ & $M$ & $R$ \\ \cline{2-5} 
 & $U$ & 1. (-1,1)  & 2. (1,-1) & 3. (0,0) \\ \cline{2-5} 
 & $C$ & 4. (1,-1) & 5. (-1,1) & 6. (0,0) \\ \cline{2-5} 
 & $D$ & 7. (0,0)  & 8. (0,0) & 9. (3,3) \\ \hline
\end{tabular}
\end{table}

Let the discrete-time strategy profile deviation process be modeled by a Markov chain, where the nine states of the Markov chain are the nine strategy profiles. At each time instant, assume the two agents will each receive a 1/2 chance of deviation and myopically choose whether to update their own strategies from the last time instant. 

As specified in the set up, an agent's decision rule is actually to choose the its transition probabilities from any strategy profile to elsewhere subject to those previously made assumptions about payoff-based deviations. An example of the transition matrix for the Markov chain, denoted by $Z$, is

\begin{equation*}
\tiny
\setlength{\tabcolsep}{4pt}
\begin{tabular}{c|ccccccccc}
            & 1.(U,L) & 2.(U,M) &  3.(U,R) & 4.(C,L)  & 5.(C,M) & 6.(C,R) & 7.(D,L) & 8.(D,M) & 9.(D,R)  \\ \hline
1.(U,L) & 1/6 & 0 & 0 & 1/2 & 0 & 0 & 1/3 & 0 & 0\\
2.(U,M) & 1/3 & 1/2 & 1/6 & 0 & 0 & 0 & 0 & 0 & 0\\
3.(U,R) & 1/2 & 0 & 0 & 0 & 0 & 0 & 0 & 0 & 1/2\\
4.(C,L) & 0 & 0 & 0 & 1/2 & 1/3 & 1/6 & 0 & 0 & 0\\
5.(C,M) & 0 & 1/3 & 0 & 0 & 1/2 & 0 & 0 & 1/6 & 0\\
6.(C,R) & 0 & 0 & 0 & 0 & 1/2 & 0 & 0 & 0 & 1/2\\
7.(D,L) & 0 & 0 & 0 & 1/2 & 0 & 0 & 0 & 0 & 1/2\\
8.(D,M) & 0 & 1/2 & 0 & 0 & 0 & 0 & 0 & 0 & 1/2\\
9.(D,R) & 0 & 0 & 0 & 0 & 0 & 0 & 0 & 0 & 1\\
\end{tabular}
\end{equation*}

Let $p^i = [p_{i}(u_1) p_{i}(u_2) ... p_{i}(u_{|\mathcal{S}|})]$ be a nonnegative $|\mathcal{S}|$-dimensional vector where the $j$-th coordinate represents the payoff of agent $i$ from playing $u_{j} \in \mathcal{S}$. The inner product of $p^i$ and the limiting distribution $\lim\limits_{n\to\infty}\frac{1}{n}\sum\limits_{k=1}^{n}\pi^{0}Z^{k}$ gives the \emph{average payoff} for agent $i$ from involving in this long-run distributed deviation process. 

From an optimization perspective, agent $i$ can choose its own individual transition probabilities to maximize its own average payoffs.  Its optimization problem can be formally stated as

 $$\max_{[\psi_{i}(u,v)]^{|\mathcal{U}| \times |\mathcal{U}|}} p^{i} \cdot \lim\limits_{n\to\infty}\frac{1}{n}\sum\limits_{k=1}^{n}\pi^{0}Z^{k},$$

%\textcolor{red}{where ... (comment: Here we need to explain the symbols. Another thing is that the multiplication of $p_i$ with the limit seems not right, as the limit is a row vector, if $p_i$ is also a row vector and you want the dot product of the two vectors, we can add a transpose to the vector in the sum like $(\pi_0 Z^k)'$.)} 

%gives an \emph{average payoff} for agent $i$ from involving in this long-run distributed deviation process.

An observation is that if at least one of the transition probabilities from the 1, 2, 4, or 5-th strategy profile to the 3, 6, 7 or 8-th strategy profile is positive (i.e., a path beginning in the cycle of the 1th, 2th, 4th and 5th strategy profiles and ending out of it exists on the strategy profile transition graph) and the initial distribution of the strategy profiles is non-degenerate (without any component to be ``1''), both agents will play the 9-th strategy profile $(D,R)$ (the only pure strategy Nash equilibrium in the game) with probability 1 in the whole process, and thus have the highest total payoffs (with 3 for both at each stage). This is the Pareto-optimal scenario, where no one can unilaterally gain without having the others worse off. If not, both agents will stay in the counter-clockwise cycle of the 1-th, 2-th, 4-th and 5-th strategy profiles, $(U,L)$, $(U,M)$, $(C,L)$, and $(C,M)$, and incur an average payoff of $0$.

The main differences of this framework in terms of equilibrium selection from the existing methods are: firstly, while predicting the limiting behavior of agents in this myopic, updating process, the distribution of equilibria that arise in this process will be predicted by the Markov chain; secondly, the prediction is done without requiring any particular interaction pattern for the agents; in other words, it can apply to the cases in which agents are connected through a network. 

More generally, how agents may distributively control or design this myopic, deviation process toward optimizing the amount of their long-term average payoffs can be studied, which thus transforms the equilibrium selection problem into an optimization problem.

\subsubsection{ From games to systems} The basic idea formalized by this approach is that the agents in the dynamical system may \emph{rationally} make their transitions based on payoffs, e.g., voters in a community network changing their opinions to increase their payoffs. One example of payoff structures is that the game is weakly acyclic. As explained in the introduction, it can reasonably be expected that this framework will give rise to different predictions than the existing conclusions derived by not assuming this payoff-based deviations.  This can be regarded as deriving predictions of interest for the dynamical system based on the game properties, which will be another possibility for future work.

%To summarize, this paper includes a framework that reformulates any finite, static game into a distributed, dynamical system of agents' distributed deviation process and suggested several applications of interest to which it can be further applied. 

%Aside from these applications, another possibility for future work is to develop theoretical extensions to cases where an agent may only know her own payoffs and study the aforementioned applications in this alternative context, such as how agents may choose their transition probabilities among strategy profiles to eventually select out equilibria of certain properties.  

\bibliographystyle{unsrt}
%\bibliography{c:/AAA/my,c:/AAA/steve,flockpaper,consensus}
\bibliography{system}

\end{document}